\documentclass[reprint,
 amsmath,amssymb,
 aps,
]{revtex4-2}

\usepackage{graphicx}
\usepackage{dcolumn}
\usepackage{bm}
\usepackage{xcolor}
\usepackage{mathptmx}
\usepackage{graphicx}
\usepackage{dcolumn}
 
\usepackage{bm}
\usepackage{amssymb}
\usepackage{comment}
\usepackage{hyperref}
\usepackage{physics}

\begin{document}

\title{Avoiding a line-of-sight obstacle via deep sub-Rayleigh shadow-projection utilizing space-time wave packets}

\author{Layton A. Hall$^{1,2}$}
\author{Bryan L. Turo$^{1}$}
\author{Ayman F. Abouraddy$^{1,*}$}

\affiliation{$^1$CREOL, The College of Optics \& Photonics, University of Central Florida, Orlando, FL 32816, USA}
\affiliation{$^2$Materials Physics and Applications - Quantum Division, Los Alamos National Laboratory, Los Alamos, NM 87545, USA}
\affiliation{$^*$raddy@creol.ucf.edu}

\begin{abstract}
A challenge in optics, which is shared by other sources of radiation, is to direct a coherent beam to impinge on a target behind an obstacle intervening in the line-of-sight (LoS). While self-accelerating or bending beams can help avoid an LoS obstacle, the beam does not reach the LoS target downstream beyond the obstacle. If one instead avoids the obstacle by projecting a \textit{transverse} null or shadow onto the axial plane at which it is located, an associated \textit{axial} shadow is cast that extends over the effective Rayleigh length, which reduces the utility of this approach. Unless either the wavelength or the transverse shadow width is changed, this Rayleigh length can only be reduced by modifying the structure of the illumination beam. Here we show that space-time wave packets (STWPs), in which each spatial frequency is tightly associated with a single wavelength, when used as an illumination beam, can dramatically reduce the axial extent of the cast shadow. Indeed, by utilizing STWPs we produce deep sub-Rayleigh-length shadows, in some cases with a more than two orders-of-magnitude reduction below the conventional Rayleigh length. For example, a 5-mm-wide transverse shadow in a Gaussian beam at a wavelength of $\sim1$~$\mu$m has a Rayleigh length of $\sim25$~m, whereas the same shadow projected by an STWP extends only $\sim0.15$~m. We demonstrate this sub-Rayleigh-length reduction in the axially cast shadow accompanying transverse nulls whose widths extend over a broad span of widths from 80~$\mu$m to 48~mm --- almost three orders-of-magnitude. These results may lead to advances in safe radiation therapy, non-LoS optical and wireless communications, selective stand-off detection, three-dimensional photolithography, and laser ablation and micro-machining. 
\end{abstract}


\maketitle

 \section{Introduction}

In a variety of settings, a laser beam (or another source of coherent radiation) is directed to an intended remote target, and it is necessary to simultaneously avoid an obstacle positioned in the line-of-sight (LoS). Such a scenario may arise in remote sensing, stand-off detection, or directed energy where a particular target is of interest while an obstacle in the LoS is to be avoided. In free-space or underwater optical or high-frequency wireless communication, the goal may be to establish connection with a receiver beyond a LoS obstacle \cite{Hu24NC,Guerboukha24CE,Chen25NC}. A further example of this scenario arises in radiation therapies where it may be needed to avoid vital or sensitive organs while delivering high-power radiation to a target tissue \cite{Dahlman83CR,Schuitmaker96JPPB,Robertson09JPPB,Schena17JFB,Chung21CR,Fan24CBS,Nestoros25NRC}. Moreover, 3D photolithography can benefit from selectively avoiding the illumination of the photoresist \cite{Jaiswal23iScience,Skliutas25NRMP,Dhand25NRB}. Similar advantages can be harnessed by laser machining, ablation, and materials processing \cite{Duocastella12LPR,Salter19LSA,Bi23JMRT,Schmidt24CIRPA}. The common feature in all of these scenarios -- although they occur over an extremely wide span of length scales -- is that the intended application is disrupted by an obstacle intervening in the LoS between the laser source and the intended target.

\begin{figure}[t!]
\centering
\includegraphics[width=8.6cm]{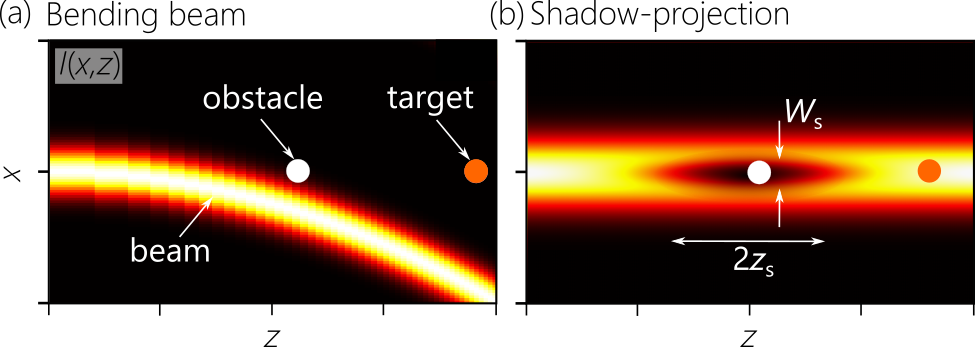}
\caption{(a) An optical beam following a curved trajectory --- whether an Airy beam or a bending STWP. Even if the bending beam avoids the LoS obstacle, it cannot reach an LoS target beyond it. (b) In shadow-projection, a null or shadow of transverse width $W_{\mathrm{s}}$ is projected onto the obstacle plane, thereby avoiding it. However, an axial shadow of total length $2z_{\mathrm{s}}$ is also cast. The target must therefore be located beyond $z_{\mathrm{s}}$ from the obstacle.}
\label{fig:Concept}
\end{figure}

A potential strategy to address this challenge is to utilize self-accelerating or bending beams that travel along curved trajectories, such as Airy beams \cite{Siviloglou07OL,Siviloglou07PRL} (or related structured beams \cite{Morris10JO,Chremmos12OL,Zhao13OL,Efremidis19Optica}), which have been used in guiding electric arcs \cite{Clerici15SciAdv} and curved plasma channels \cite{Polynkin09Science}. Recently, spatiotemporally structured pulsed beams in the form of space-time wave packets (STWPs) \cite{Kondakci16OE,Parker16OE,Kondakci17NP,Yessenov22NC,Yessenov22AOP} (in which each spatial frequency is tightly associated with a prescribed wavelength) have been utilized to synthesize bending beams \cite{Hall25OLbending} (studied also in Refs.~\cite{Liang23OL,Wang25OL}). However, such curved-trajectory beams do not avoid an LoS obstacle while simultaneously maintaining incidence on an LoS target lying downstream beyond the LoS obstacle [Fig.~\ref{fig:Concept}(a)]. An alternative strategy to address this challenge is to project a shadow onto the obstacle. By introducing a transverse null of width larger than that of the obstacle into the beam and then relaying it to the plane of the LoS obstacle, no radiation impinges on it [Fig.~\ref{fig:Concept}(b)]. However, inevitably, a concomitant \textit{axial} shadow is cast, which extends for the associated Rayleigh length. For example, when using a collimated field at a wavelength $\lambda_{\mathrm{o}}\approx1$~$\mu$m, a shadow of transverse width $W_{\mathrm{s}}\approx5$~mm casts an axial shadow that extends for $>\tfrac{W_{\mathrm{s}}^{2}}{\lambda_{\mathrm{o}}}\sim25$~m \cite{SalehBook07}. Consequently, one cannot arrange for a conventional laser beam to impinge on a target while avoiding an LoS obstacle in its close proximity. Except for reducing the transverse width of the projected shadow or increasing the wavelength, the only alternative to change the length of the axial cast-shadow is to modify the structure of the illumination field itself.

Here we show that utilizing an STWP as an illumination field can help reduce the axial length of the cast shadow to deep sub-Rayleigh lengths, thereby offering a unique advantage for shadow-projection in applications that require avoiding an LoS obstacle while maintaining incidence on a close LoS target, thus providing more precise discrimination between contiguous targets. By introducing a transverse spatial null into the STWP at the source and relaying the field to the obstacle, the spatiotemporally structured field avoids the LoS obstacle and then rapidly returns to its original form. This takes place over an axial distance that is orders-of-magnitude smaller compared to that accompanying a shadow of equal transverse width projected by a conventional beam at the same wavelength. We show that any illumination beam for such an application must satisfy two desiderata: the transverse field width must exceed the width of the obstacle, and yet the spatial bandwidth of the field must simultaneously exceed that of the obstacle. These seemingly incompatible requirements are realized using STWPs. Because each spatial frequency in an STWP is tightly associated with a prescribed wavelength \cite{Kondakci16OE,Parker16OE}, the \textit{time-averaged intensity} profile is rendered \textit{effectively incoherent spatially} (despite absence of any statistical fluctuations). The purely spatial coherence function for the STWP in the spectral domain -- which governs the propagation dynamics of the time-averaged spatial profile and thus the axial length of the cast shadow -- has two independent scales: the full spatial bandwidth that can be broad and which determines the axial shadow length, and a much narrower spatial uncertainty \cite{Kondakci19OL,Yessenov19OE} that determines the transverse spatial width of the STWP and thus the size of the obstacle that can be avoided. Critically, the field retrieves its form along the optical axis in absence of the obstacle after this sub-Rayleigh-length axial shadow. In laboratory-scale table-top experiments, we avoid LoS obstacles of widths ranging from 80~$\mu$m to 5~mm, and in experiments outside the laboratory down a corridor we avoid obstacles of width extending up to 48~mm. In all cases, we realize reduction factors of the axial shadow extending from $\approx25\times$ to $\approx250\times$ with respect to the effective Rayleigh length. The measurement results are in agreement with a theoretical model that takes into account the dual length scales intrinsically characterizing STWPs. Furthermore, we discuss other potential optical field configurations that satisfy the desiderata for sub-Rayleigh-length shadow-projection. These results suggest a host of applications in remote sensing, photodynamic radiation therapies, directed energy, laser machining, three-dimensional photolithography, and free-space optical and millimeter-wave communications.

\begin{figure}[t!]
\centering
\includegraphics[width=8.6cm]{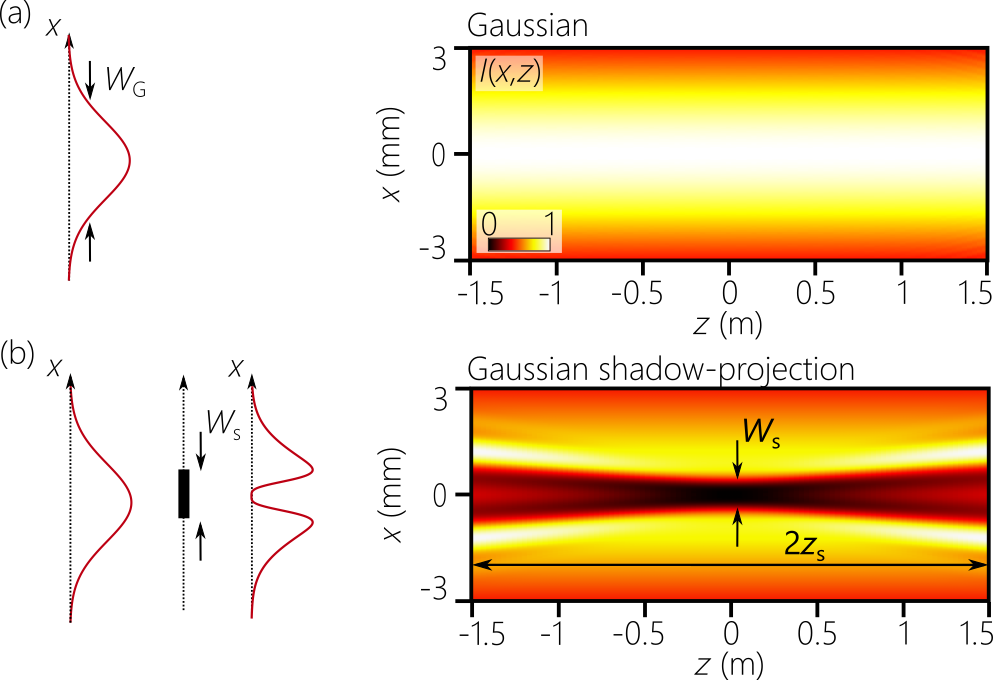}
\caption{(a) A Gaussian beam of width $W_{\mathrm{G}}=6$~mm at $\lambda_{\mathrm{o}}\sim1$~$\mu$m (left), and the intensity profile $I(x,z)$ (right). This beam has a Rayleigh length $z_{\mathrm{G}}\approx30$~m, so there is no variation in $I(x,z)$ over the axial range plotted. (b) A portion of the Gaussian beam in (a) of width $W_{\mathrm{s}}=1.2$~mm is blocked (left). The beam is relayed to the mid-range of $I(x,z)$ (right). A transverse null of width $W_{\mathrm{s}}$, accompanied by an axially cast shadow of Rayleigh length $z_{\mathrm{G,s}}\approx1.44$~m.}
\label{fig:GaussianShadow}
\end{figure}

\section{Line-of-sight obstacle-avoidance using a Gaussian beam}

\subsection{Shadow-projection using a Gaussian beam}

A Gaussian beam of width $W_{\mathrm{G}}$ at a wavelength $\lambda_{\mathrm{o}}$ has a Rayleigh length $z_{\mathrm{G}}\sim\tfrac{W_{\mathrm{G}}^{2}}{\lambda_{\mathrm{o}}}$ as shown in the intensity profile $I(x,z)$ [Fig.~\ref{fig:GaussianShadow}(a)]; here $z$ is the axial coordinate and $x$ is the transverse coordinate (we hold the field uniform along $y$; we return to the case of two transverse coordinates in the Discussion). To cast a shadow, we block a portion of the Gaussian beam at the source and then image the beam to the axial plane where an obstacle of width $W_{\mathrm{s}}$ is located [Fig.~\ref{fig:GaussianShadow}(b)]. The transverse width of the beam must of course be larger than the obstacle ($W_{\mathrm{G}}>W_{\mathrm{s}}$), or else the beam is entirely eliminated by the block. A concomitant axial null is cast accompanying this transverse null, which extends for a length $\pm z_{\mathrm{G,s}}$ around the central shadow plane, where: 
\begin{equation}\label{eq:ShadowRayleigh}
z_{\mathrm{G,s}}\sim\frac{W_{\mathrm{s}}^{2}}{\lambda_{\mathrm{o}}}.
\end{equation}
A single transverse spatial scale ($W_{\mathrm{G}}$ or $W_{\mathrm{s}}$) contributes significantly to the determination of the axial scale ($z_{\mathrm{G}}$ or $z_{\mathrm{G,s}}$, respectively). Note that the width of the Gaussian beam $W_{\mathrm{G}}$ does not impact the axial shadow length significantly as long as $W_{\mathrm{G}}$ is larger than $W_{\mathrm{s}}$ (typically when $W_{\mathrm{G}}>2W_{\mathrm{s}}$). If instead $W_{\mathrm{G}}\rightarrow W_{\mathrm{s}}$, then additional diffractive effects mar the shadow formation. Note further that $z_{\mathrm{G,s}}$ is only a crude and quite conservative estimate for the axial shadow length. At $z_{\mathrm{G,s}}$ beyond the shadow center, the peak intensity is not yet on the optical axis (the shadow has not `healed').  
 
Our goal here is to discriminate between an LoS obstacle of width $W_{\mathrm{s}}$ and an intended target that is in the LoS downstream beyond the obstacle, so that the beam impinges on the target while avoiding the LoS obstacle. If the target lies within the axially cast shadow, it also does not receive optical radiation. We thus aim to reduce the length of the axial shadow cast below the Rayleigh-length $W_{\mathrm{G,s}}\sim W_{\mathrm{s}}^{2}/\lambda_{\mathrm{o}}$ while holding $W_{\mathrm{s}}$ and $\lambda_{\mathrm{o}}$ fixed by modifying the \textit{structure} of the illumination field. 

\begin{figure}[t!]
\centering
\includegraphics[width=8.6cm]{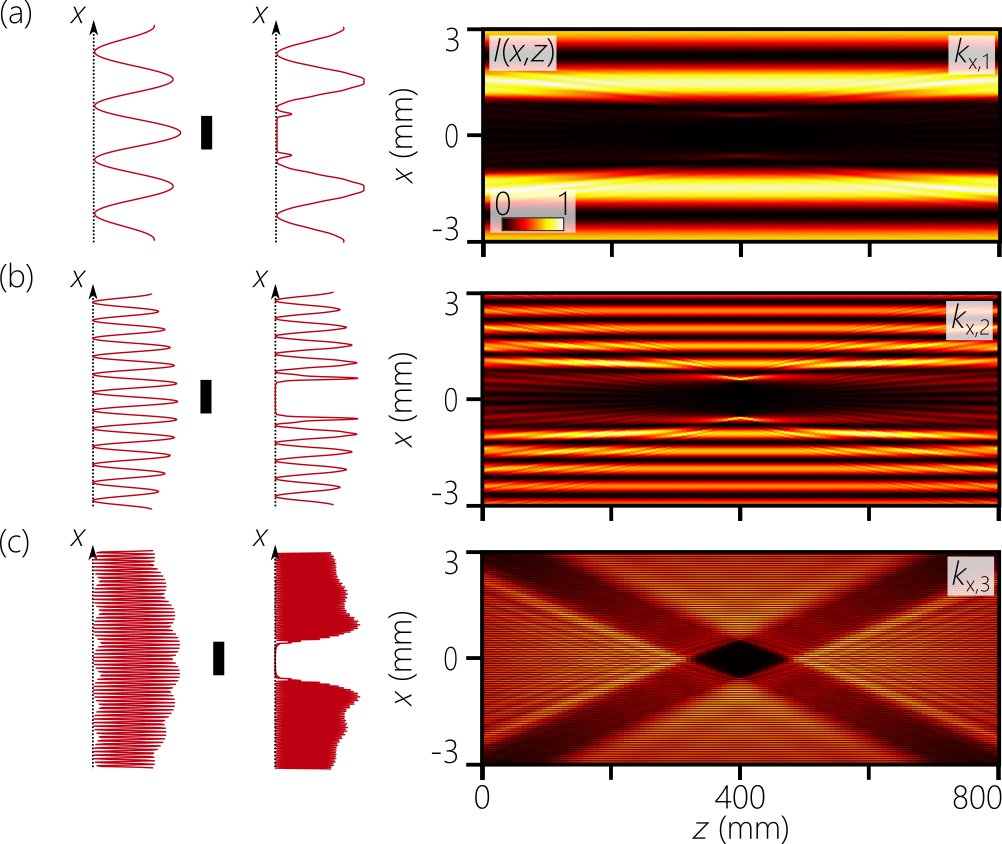}
\caption{(a-c) An extended cosine wave $\cos(k_{x}x)$ traverses an obstruction of width $W_{\mathrm{s}}=1$~mm. On the right we plot the intensity profile $I(x,z)$. The obstructed cosine wave is relayed to the plane $z=400$~mm. (a) A low spatial frequency $k_{x1}<\tfrac{2\pi}{W_{\mathrm{s}}}$, where less than a single period of the cosine wave is blocked; (b) an intermediate spatial frequency $k_{x2}=\tfrac{2\pi}{W_{\mathrm{s}}}$, where a full period of the cosine wave is blocked; and (c) a high spatial frequency $k_{x3}>\tfrac{2\pi}{W_{\mathrm{s}}}$, where multiple periods of the cosine wave are blocked. The axial length of the cast shadow drops with increasing $k_{x}$.}
\label{fig:Coherent}
\end{figure}

\subsection{Rayleigh-length shadow-projection: coherent superposition}

To elucidate the requirements for sub-Rayleigh-length shadow-projection, we examine in Fig.~\ref{fig:Coherent}(a) the shadow cast by relaying an extended cosine wave $\cos(k_{x}x)$ (formed of a pair of plane waves of the same wavelength symmetrically tilted with the $z$-axis) after blocking a portion of the beam of width $W_{\mathrm{s}}$. We write the spatial frequency as $k_{x}=\tfrac{2\pi}{\lambda_{\mathrm{o}}}\sin\varphi$, where $\varphi$ is the tilt angle of the wave vector with the $z$-axis, and $\sin\varphi\approx\varphi$ in the paraxial regime. Casting a shadow of transverse width $W_{\mathrm{s}}$ is associated with an axial shadow of length estimated geometrically as $z_{\mathrm{s}}\sim\tfrac{W_{\mathrm{s}}}{\varphi}$.

\begin{figure}[t!]
\centering
\includegraphics[width=8.6cm]{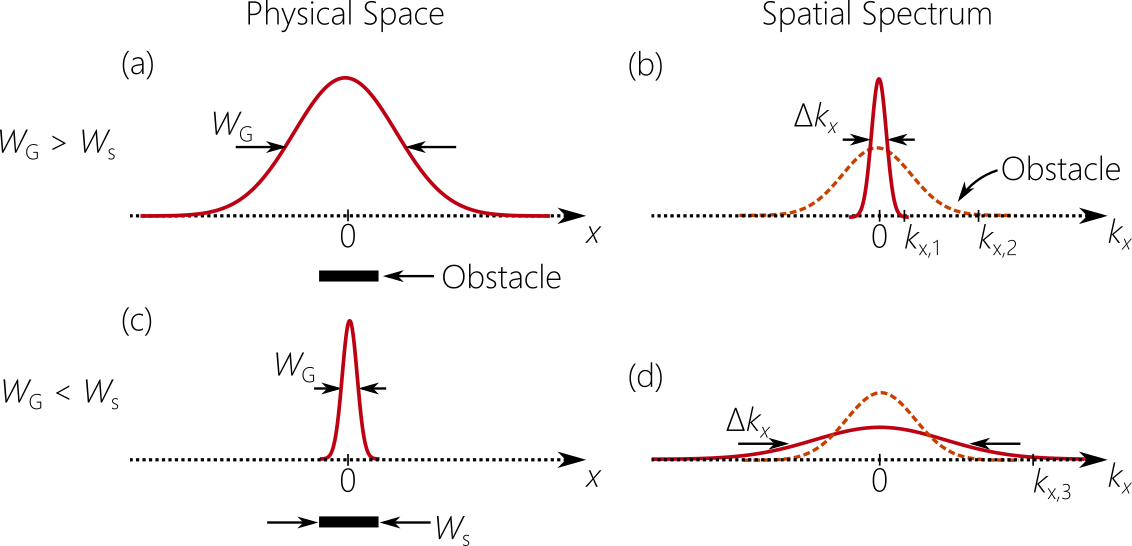}
\caption{(a) A Gaussian beam of width $W_{\mathrm{G}}$ is used to cast a shadow of width $W_{\mathrm{s}}<W_{\mathrm{G}}$. (b) The spatial spectrum of the Gaussian beam in (a) is narrower than that of the obstacle. The spatial bandwidth of the Gaussian beam is $k_{x1}$ [Fig.~\ref{fig:Coherent}(a)], and that of the obstacle is $k_{x2}$ [Fig.~\ref{fig:Coherent}(b)]. (c) If the width of the Gaussian beam $W_{\mathrm{G}}$ is made smaller to reduce the associated Rayleigh length of the axially cast shadow, we reach the limit of $W_{\mathrm{G}}\sim W_{\mathrm{s}}$, and the beam is blocked. (d) The spatial spectrum of the Gaussian beam, of bandwidth $k_{x3}$ [Fig.~\ref{fig:Coherent}(c)], is now wider than that of the obstacle.}
\label{fig:BeamSize}
\end{figure}

When $k_{x}$ is small [$k_{x1}$ in Fig.~\ref{fig:Coherent}(a)], corresponding to a small tilt angle $\varphi_{1}$ with the $z$-axis, a long axial shadow is cast. Here, less than one period of the cosine wave is blocked when $k_{x1}<\tfrac{\pi}{W_{\mathrm{s}}}$, whereupon $z_{\mathrm{s}1}\sim\tfrac{W_{\mathrm{s}}}{\lambda_{\mathrm{o}}}\tfrac{\pi}{k_{x1}}>\tfrac{W_{\mathrm{s}}^{2}}{\lambda_{\mathrm{o}}}$ -- corresponding to a \textit{super}-Rayleigh-length projected shadow. An intermediate spatial frequency $k_{x2}\approx\tfrac{2\pi}{\lambda_{\mathrm{o}}}\varphi_{2}$ (larger tilt angle with the $z$-axis) casts as expected a shorter axial shadow associated with a transverse shadow of the same width [Fig.~\ref{fig:Coherent}(b)]. A single period of the cosine wave is blocked when $k_{x2}=\tfrac{\pi}{W_{\mathrm{s}}}$, whereupon the axial shadow has a length $z_{\mathrm{s}2}\sim\tfrac{W_{\mathrm{s}}}{\varphi_{2}}\sim\tfrac{W_{\mathrm{s}}^{2}}{\lambda_{\mathrm{o}}}$, which is the Rayleigh length associated with the transverse width $W_{\mathrm{s}}$. Finally, when a large spatial frequency $k_{x3}=\tfrac{2\pi}{\lambda_{\mathrm{o}}}\varphi_{3}>\tfrac{\pi}{W_{\mathrm{s}}}$ is used [Fig.~\ref{fig:Coherent}(c)], multiple periods of the cosine wave are blocked. The axial shadow has a length $z_{\mathrm{s}3}\sim\tfrac{W_{\mathrm{s}}}{\varphi_{3}}<\tfrac{W_{\mathrm{s}}^{2}}{\lambda_{\mathrm{o}}}$, corresponding to a \textit{sub}-Rayleigh-length projected shadow.

It is clear from Fig.~\ref{fig:Coherent} that higher spatial frequencies contribute to reducing the length of the axially cast shadow, and it is thus advantageous from this perspective to increase the spatial bandwidth $\Delta k_{x}$ of the optical source to include more high spatial frequencies when used in shadow-projection. However, Fig.~\ref{fig:BeamSize} indicates that taking this route with a Gaussian beam (or other conventional beams) is not feasible. A Gaussian beam is a coherent superposition of cosine waves limited by the spatial bandwidth $\Delta k_{x}\sim\tfrac{\pi}{W_{\mathrm{G}}}$ having a Rayleigh length $z_{\mathrm{G}}\sim\tfrac{W_{\mathrm{G}}^{2}}{\lambda_{\mathrm{o}}}$, whereas the transverse shadow width is $W_{\mathrm{s}}$ associated with a spatial bandwidth $\sim\tfrac{\pi}{W_{\mathrm{s}}}$ and Rayleigh length $\sim\tfrac{W_{\mathrm{s}}^{2}}{\lambda_{\mathrm{o}}}$ [Fig.~\ref{fig:BeamSize}(a,b)]. Increasing the spatial bandwidth of the Gaussian beam to include higher spatial frequencies (and thus reducing the length of the axially cast shadow) accompanies a \textit{reduction} in the beam width $W_{\mathrm{G}}$, which is limited by the requirement that $W_{\mathrm{G}}>W_{\mathrm{s}}$ -- or else the beam cannot support projecting a shadow of width $W_{\mathrm{s}}$ [Fig.~\ref{fig:BeamSize}(c,d)]. In other words, these two requirements are mutually exclusive in conventional beams: reducing the axial Rayleigh length while maintaining a large transverse width.

\subsection{Criterion for an illumination field to yield sub-Rayleigh-length shadow-projection}

As shown above, conventional illumination fields are limited to Rayleigh-length shadow-projection. We can formulate the criterion for the illumination field to yield \textit{sub}-Rayleigh-length shadow-projection as follows:
\begin{enumerate}
\item The transverse extent of the illumination field $W$ must exceed the transverse width $W_{\mathrm{s}}$ of the obstacle to be avoided  (the width of the projected shadow), $W>W_{\mathrm{s}}$ [Fig.~\ref{fig:BeamSize}(a,c)].
\item The beam spatial bandwidth $\Delta k_{x}$ must be large to incorporate high spatial frequencies necessary for reducing the axial length of the cast shadow; specifically, $\Delta k_{x}>\tfrac{\pi}{W_{\mathrm{s}}}$ [Fig.~\ref{fig:Coherent}(c) and Fig.~\ref{fig:BeamSize}(b,d)]. 
\end{enumerate}

These two desiderata are incompatible when using a conventional coherent beam in which the beam width $W$ is related directly to its spatial bandwidth $\Delta k_{x}\sim\tfrac{\pi}{W}$; increasing the spatial bandwidth is associated with a reduction in axial shadow length. We proceed to show that STWPs offer a way to combine these two desiderata by severing the link between the beam width and its spatial bandwidth.

\section{Line-of-sight obstacle-avoidance using a space-time wave packet}

\subsection{STWPs}

\begin{figure}[t!]
\centering
\includegraphics[width=8.6cm]{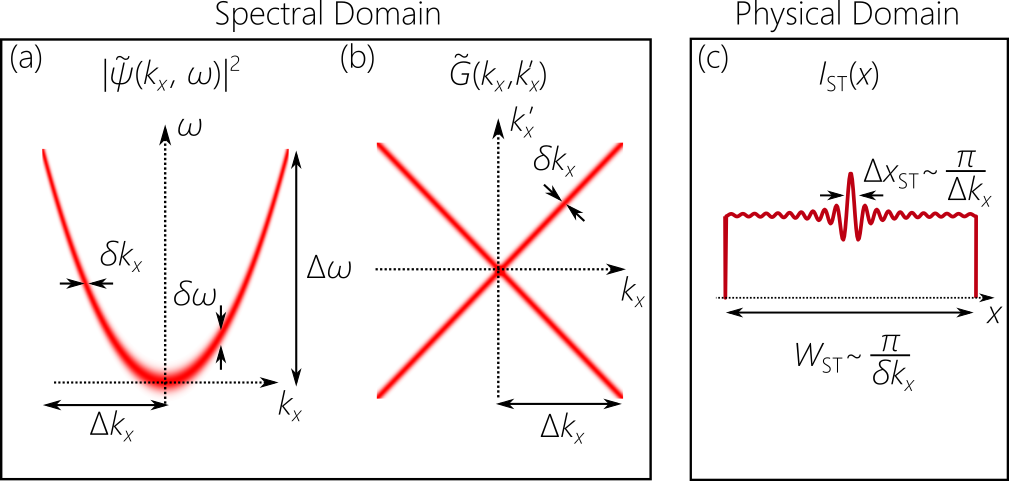}
\caption{(a) The spatiotemporal spectrum of a typical STWP $|\widetilde{\psi}(k_{x},\omega)|^{2}$. The spatial bandwidth is $\Delta k_{x}$ and the temporal bandwidth is $\Delta\omega$. The association between spatial and temporal frequencies is limited by an uncertainty that can be expressed as $\delta\omega$ or $\delta k_{x}$. (b) From the spatiotemporal spectrum in (a), we obtain the purely spatial coherence spectral function $\widetilde{G}(k_{x},k_{x}')$ after tracing over $\omega$, highlighting the two spatial bandwidths: the overall spatial bandwidth $\Delta k_{x}$ and the spatial uncertainty $\delta k_{x}$. This double-branched spatial coherence function governs the propagation of the time-averaged intensity $I(x,z)$ of the STWP. (c) The transverse spatial intensity profile $I(x,z=0)$ highlighting the two STWP spatial scales: the outer width $W_{\mathrm{ST}}\sim\tfrac{\pi}{\delta k_{x}}$ and width of the central spatial feature $\Delta x_{\mathrm{ST}}\sim\tfrac{\pi}{\Delta k_{x}}$.}
\label{fig:CoherenceFunction}
\end{figure}

An STWP is a pulsed beam in which each constitutive plane wave (associated with a transverse wave number or spatial frequency $\pm k_{x}$) is associated with a single temporal frequency $\omega$ (or wavelength $\lambda$). We consider here STWPs localized along one transverse dimension $x$ while remaining uniform along the other (we return to the 2D scenario in the Discussion). A propagation-invariant STWP travels rigidly in free space when the axial wave number $k_{z}$ is linear in the temporal frequency $\omega$, $\Omega=(k_{z}-k_{\mathrm{o}})c\tan\theta$, where $\Omega=\omega-\omega_{\mathrm{o}}$, $\omega_{\mathrm{o}}$ is a fixed temporal frequency, $k_{\mathrm{o}}=\omega_{\mathrm{o}}/c$, $c$ is the speed of light in vacuum. This constraint in the $(k_{x},k_{z},\tfrac{\omega}{c})$-space corresponds to a plane that is parallel to the $k_{x}$-axis and makes an angle $\theta$ (the spectral tilt angle) with the $k_{z}$-axis. Combining this constraint with the free-space dispersion relationship $k_{x}^{2}+k_{z}^{2}=(\tfrac{\omega}{c})^{2}$, we obtain:
\begin{equation}\label{eq:Parabola}
\frac{\Omega(k_{x})}{\omega_{\mathrm{o}}}\approx\frac{k_{x}^{2}}{2k_{\mathrm{o}}^{2}(1-\cot\theta)};
\end{equation}
see Fig.~\ref{fig:CoherenceFunction}(a). The STWP approaches a pulsed plane-wave when $\theta\rightarrow45^{\circ}$, and a monochromatic beam when $\theta\rightarrow0^{\circ}$.

The spatiotemporal envelope of the STWP field $E(x,z;t)=e^{i(k_{\mathrm{o}}z-\omega_{\mathrm{o}}t)}\psi(x,z;t)$ is expressed as an angular spectrum: $\psi(x,z;t)\!=\!\int\!dk_{x}\widetilde{\psi}(k_{x})e^{i\{k_{x}x+(k_{z}-k_{\mathrm{o}})z-\Omega(k_{x})t\}}\!=\!\psi(x,0;t-\tfrac{z}{c\tan\theta})$ \cite{Kondakci19NC}. The time-averaged intensity $I(x,z)\!=\!\int\!dt|E(x,z;t)|^{2}=I(x,0)$ is diffraction-free [Fig.~\ref{fig:STWPShadow}(a)]:
\begin{equation}\label{eq:TimeAveraged}
I(x,z)=\int\!dk_{x}|\widetilde{\psi}(k_{x})|^{2}+\int\!dk_{x}|\widetilde{\psi}(k_{x})|^{2}\cos(2k_{x}x);
\end{equation}
which is formed of mutually incoherent cosine waves; we have assumed that $\widetilde{\psi}(k_{x})$ is an even function without loss of generality \cite{Yessenov19OE,Yessenov19Optica}. For a spatial bandwidth $\Delta k_{x}$, the time-averaged intensity profile comprises a flat pedestal atop of which is formed a localized spatial feature of transverse width $\Delta x\sim\tfrac{\pi}{\Delta k_{x}}$ [Fig.~\ref{fig:CoherenceFunction}(c)]. Because of the factor-of-2 in the cosine term in Eq.~\ref{eq:TimeAveraged}, the width $\Delta x$ here is half that of a coherent superposition of cosines of the same spatial bandwidth $\Delta k_{x}$ \cite{Yessenov19Optica}. This result (the apparent incoherence as a result of tracing out a degree-of-freedom of the field that is tightly correlated to another) is a well-known consequence of what has come to be known as `classical entanglement' \cite{Kagalwala13NP,Kondakci19OL}. In the case examined here, starting with a coherent field in which a tight association exists between the spatial and temporal frequencies, tracing out the temporal spectrum (corresponding to the time-averaged intensity) yields a spatial coherence function that effectively represents a partially coherent field. 

The ideal scenario of an exact delta-function association between axial wave number $k_{z}$ and temporal frequency $\omega$ (and thus, in turn, between $\pm k_{x}$ and $\omega$; Eq.~\ref{eq:Parabola}) cannot be attained in practice because it requires infinite energy \cite{Sezginer85JAP}. Instead, finite resources impose a spectral uncertainty $\delta\omega$ or equivalently a spatial uncertainty $\delta k_{x}$ existing between $k_{x}$ and $\omega$ [Fig.~\ref{fig:CoherenceFunction}(a)]. For spectral tilt angles $\theta\rightarrow45^{\circ}$ and an aperture of width $W_{\mathrm{ST}}$ for the STWP, the accompanying uncertainty in the spatial spectrum is
$\delta k_{x}\sim\tfrac{\pi}{W_{\mathrm{ST}}}$, which limits the extent over which the pedestal in $I(x,z)$ extends laterally. The time-averaged intensity in this case is given by:
\begin{equation}
I(x,z)=\iint\!dk_{x}dk_{x}'\widetilde{G}(k_{x},k_{x}')e^{i(k_{x}-k_{x}')x}e^{-i\left(\tfrac{k_{x}^{2}-k_{x}'^{2}}{2k_{\mathrm{o}}}\right)z},
\end{equation}
where $\widetilde{G}(k_{x},k_{x}')$ is a correlation function given by:
\begin{equation}
\widetilde{G}(k_{x},k_{x}')=\int\!d\Omega\;\widetilde{\psi}(k_{x},\Omega)\widetilde{\psi}^{*}(k_{x}',\Omega).
\end{equation}
For the parabolic spatiotemporal spectrum $\widetilde{\psi}(k_{x},\Omega)$ shown in Fig.~\ref{fig:CoherenceFunction}(a) (Eq.~\ref{eq:Parabola}), the corresponding correlation function is shown in Fig.~\ref{fig:CoherenceFunction}(b), which resembles a double-branched (symmetric/antisymmetric) Gaussian-Schell model. The total spectral bandwidth $\Delta k_{x}$ determines the width $\Delta x_{\mathrm{ST}}\sim\tfrac{\pi}{\Delta k_{x}}$ of the central spatial feature in $I(x,0)$ [Fig.~\ref{fig:CoherenceFunction}(c)], whereas the width $\delta k_{x}$ of each branch in $\widetilde{G}(k_{x},k_{x}')$ determines the outer width of the STWP: $W_{\mathrm{ST}}\sim\tfrac{\pi}{\delta k_{x}}$. The propagation dynamics of such an STWP is shown in Fig.~\ref{fig:STWPShadow}(a).

In contrast to conventional optical beams that are characterized by a single transverse length scale $W_{\mathrm{G}}$, an STWP is characterized by two \textit{independent} transverse length scales: $W_{\mathrm{ST}}$ and $\Delta x_{\mathrm{ST}}$. Typically $W_{\mathrm{ST}}\gg\Delta x_{\mathrm{ST}}$, which reflects the fact that the uncertainty $\delta k_{x}\sim\tfrac{\pi}{W_{\mathrm{ST}}}$ is typically much smaller than the spatial bandwidth $\Delta k_{x}\sim\tfrac{\pi}{\Delta x_{\mathrm{ST}}}$. It can be shown that when $\theta\rightarrow45^{\circ}$, the diffraction-free length for an STWP is \cite{Hall25OE1km,Hall25OL}:
\begin{equation}\label{eq:DistanceSTWP}
z_{\mathrm{ST}}\sim\frac{W_{\mathrm{ST}}\Delta x_{\mathrm{ST}}}{\lambda_{\mathrm{o}}};
\end{equation}
a different formula applies when $\theta$ deviates significantly from $45^{\circ}$ \cite{Yessenov19OE}. It is critical to compare the propagation distance $z_{\mathrm{ST}}$ for an STWP (Eq.~\ref{eq:DistanceSTWP}) to the Rayleigh length $z_{\mathrm{G}}$ for a conventional beam at the same wavelength. Both propagation lengths have the same dependence on $\lambda_{\mathrm{o}}$. However, $z_{\mathrm{G}}$ depends quadratically on $W_{\mathrm{G}}$, whereas $z_{\mathrm{ST}}$ depends instead on the product of the two disparate length scales $W_{\mathrm{ST}}$ and $\Delta x_{\mathrm{ST}}$. If we take $W_{\mathrm{ST}}\sim W_{\mathrm{G}}$ (i.e., similar apertures), then $z_{\mathrm{ST}}<z_{\mathrm{G}}$. However, if we take a conventional beam whose width matches that of the central feature of an STWP, $W_{\mathrm{G}}\sim\Delta x_{\mathrm{ST}}$, then $z_{\mathrm{G}}<z_{\mathrm{ST}}$.

\subsection{Shadow-projection with an STWP}

An STWP thus satisfies both desiderata for sub-Rayleigh-length shadow-projection: (1) its outer width $W_{\mathrm{ST}}$ can be large (and can be increased further by reducing the uncertainty $\delta k_{x}$), which helps accommodate large-sized obstacles; and (2) its spatial bandwidth $\Delta k_{x}$ \textit{can also be large}, which helps reduce the effective Rayleigh length of the axially cast shadow. This is possible because the two quantities $W_{\mathrm{ST}}$ and $\Delta k_{x}$ are unrelated, in contrast to conventional beams in which these two quantities are inversely proportional to each other.

\begin{figure}[t!]
\centering
\includegraphics[width=8.6cm]{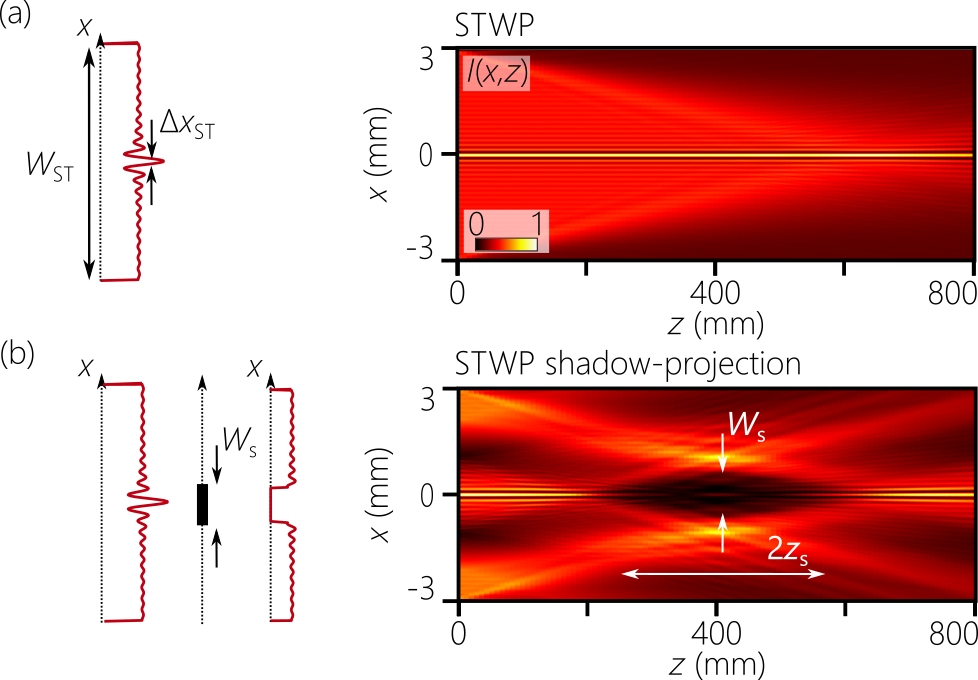}
\caption{(a) A propagation-invariant STWP whose time-averaged intensity profile is formed of a central spatial feature of width $\Delta x_{\mathrm{ST}}=150$~$\mu$m atop a pedestal of width $W_{\mathrm{ST}}=6$~mm. On the right, we plot the time-averaged intensity profile $I(x,z)$; here $\Delta\lambda=25$~nm and $\lambda_\mathrm{o}=1075$~nm. The diffraction-free length of the STWP is $z_{\mathrm{ST}}\approx 900$~mm. (b) The STWP in (a) traverses a block of width $W_{\mathrm{s}}=1$~mm (left), and this plane is relayed to $z=400$~mm, as seen in the intensity profile $I(x,z)$ (right). A transverse null of width $W_{\mathrm{s}}$ is formed, accompanied by an axially cast shadow of length $2z_{\mathrm{ST,s}}$, where $z_{\mathrm{ST,s}}\approx 150$~mm. Note that $z_{\mathrm{ST,s}}\ll z_{\mathrm{s}}$ for the axial shadow of the same transverse width cast by a Gaussian beam in Fig.~\ref{fig:GaussianShadow}(b).}
\label{fig:STWPShadow}
\end{figure}

Consider now the shadow cast by an STWP after blocking a portion of width $W_{\mathrm{s}}<W_{\mathrm{ST}}$ and then relaying the beam to the obstacle plane. We plot in Fig.~\ref{fig:STWPShadow}(b) the calculated intensity profile $I(x,z)$, where we observe a shadow of transverse width $W_{\mathrm{s}}$ and axial length $\pm z_{\mathrm{ST,s}}$, where:
\begin{equation}
z_{\mathrm{ST,s}}\sim\frac{W_{\mathrm{s}}\Delta x_{\mathrm{ST}}}{\lambda_{\mathrm{o}}};    
\end{equation}
where $\Delta x_{\mathrm{ST}}$ replaces $W_{\mathrm{s}}$ in Eq.~\ref{eq:ShadowRayleigh}. If the STWP has $\Delta x_{\mathrm{ST}}\ll W_{\mathrm{s}}$ (the width of the STWP central feature is smaller than the width of the object to be avoided), then the axial shadow length $z_{\mathrm{ST,s}}\ll z_{\mathrm{G,s}}$ [compare Fig.~\ref{fig:STWPShadow}(b) to Fig.~\ref{fig:GaussianShadow}(b)], which we refer to as sub-Rayleigh shadow-projection. Indeed, the reduction factor in the axial shadow length is $\tfrac{W_{\mathrm{s}}}{\Delta x_{\mathrm{ST}}}$, which allows for the separation between the intended target and unintended obstacle to be much smaller when relying on an STWP as an illumination field [Fig.~\ref{fig:STWPShadow}(b)] than when utilizing a conventional beam [Fig.~\ref{fig:GaussianShadow}(b)].

For spectral tilt angles $\theta$ that are close to $45^{\circ}$, $\theta=45^{\circ}+\delta\theta$, then we have $\tfrac{\Delta x_{\mathrm{ST}}}{\lambda_{\mathrm{o}}}=\tfrac{1}{2\sqrt{\delta\theta\Delta\lambda/\lambda_{\mathrm{o}}}}$, with $\delta\theta$ in radians. The reduction factor of the axial cast shadow is :
\begin{equation}
\eta_{\mathrm{ST,G}}=\frac{z_{\mathrm{s}}}{z_{\mathrm{ST,s}}}\sim2\frac{W_{\mathrm{s}}}{\lambda_{\mathrm{o}}}\sqrt{\delta\theta\frac{\Delta\lambda}{\lambda_{\mathrm{o}}}}.
\end{equation}
Therefore, the axial length of the cast shadow can be reduced while avoiding a target of \textit{fixed} transverse width by increasing $\delta\theta$ (the spectral tilt angle deviates further from $45^{\circ}$) or increasing the temporal bandwidth utilized. 

\section{Experimental setup}

\begin{figure}[t!]
\centering

\includegraphics[width=8.6cm]{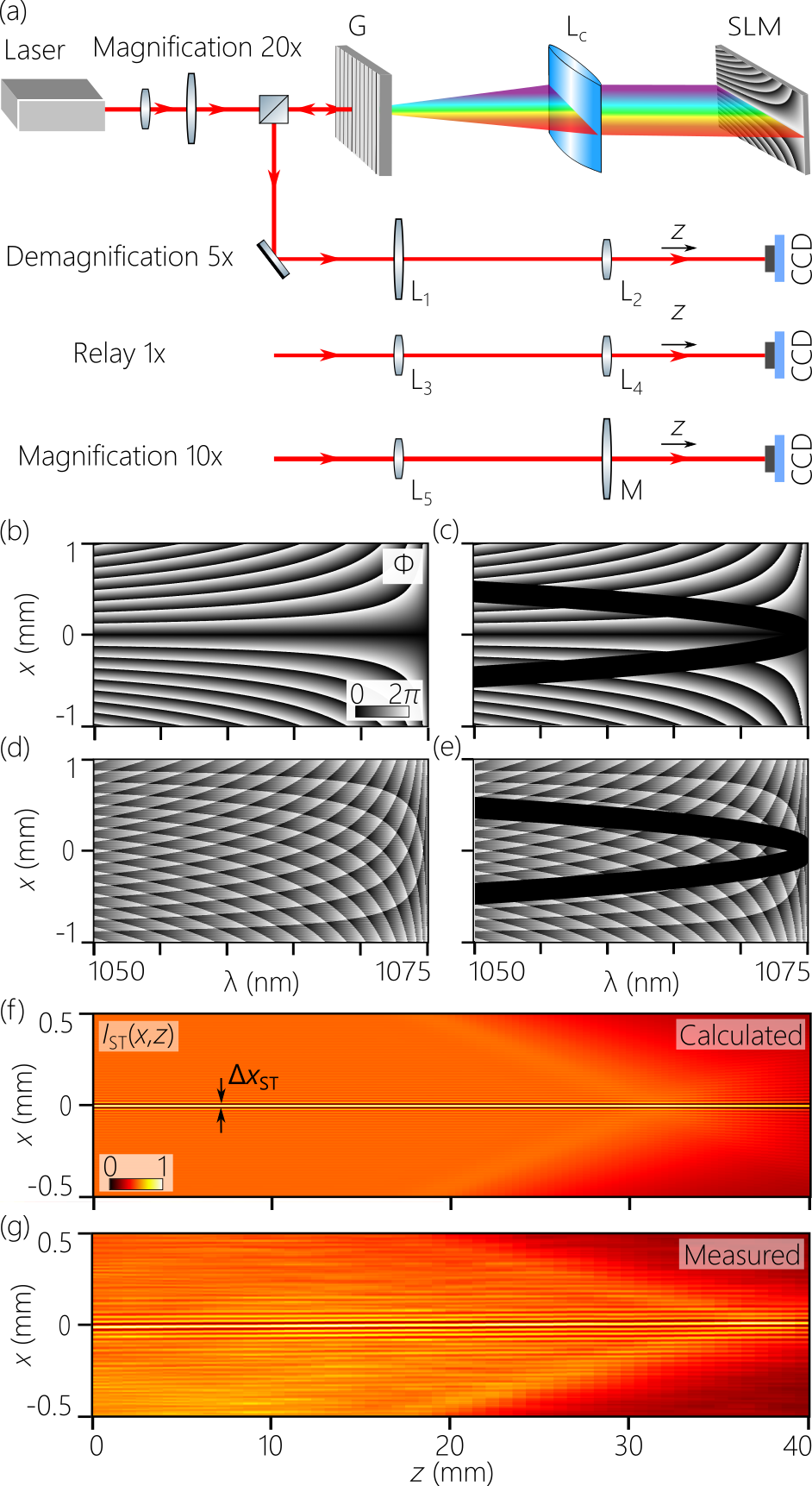}
\caption{(a) Setup for synthesizing STWPs. G: Diffraction grating; L$_{\mathrm{c}}$: cylindrical lens; SLM: spatial light modulator. The STWP is relayed by one of three optical imaging systems: a two-lens $\times5$ demagnification system (L$_{1}$ and L$_{2}$); a two-lens $1\times$ relay system (L$_{3}$ and L$_{4}$); or a $\times10$ magnification system formed of a lens L$_{5}$ and a large-aperture curved mirror~M. (b-e) Examples of the SLM phase distributions $\Phi$. (b) $\Phi$ for a propagation-invariant STWP at $\theta=45.05^{\circ}$, and (c) for the STWP in (b) after introducing a shadow of transverse width 0.2~mm. The lower and upper halves of the phase distribution correspond to positive and negative valued spatial frequencies, respectively. (d,e) Same as (b,c) after \textit{interleaving} the positive and negative spatial frequencies in adjacent columns. (f) Calculated time-averaged intensity $I(x,z)$ for an STWP produced by an interleaved SLM phase distribution having $\Delta x_{\mathrm{ST}}=12.4$~$\mu$m, $\Delta\lambda=25$~nm, and $\lambda_{\mathrm{o}}=1075$~nm. (g) Measured time-averaged intensity $I(x,z)$ corresponding to the parameters in (f).}
\label{fig:Setup}
\end{figure}

The experimental arrangement is depicted in Fig.~\ref{fig:Setup}(a), which is based on the universal angular-dispersion synthesizer in Refs.~\cite{Hall24JOSAA,Romer25JO}. We make use of laser pulses from a mode-locked laser (Spark Laser, Alcor) at a central wavelength $\approx1064$~nm, width $\approx100$~fs, bandwidth $\approx25$~nm, and 80-MHz repetition rate. The laser beam width of 2~mm is expanded to 40~mm before impinging on a diffraction grating (1200~lines/mm, $50\times50$~mm$^{2}$) at an angle $24^{\circ}$. The first diffraction order is selected and collimated via a cylindrical lens (focal length $f=250$~mm, aperture 25~mm) to spatially resolve the temporal spectrum before impinging on a reflective SLM (Meadowlark, E19X12) placed at the focal plane of the lens. The SLM has an active area of $15.36\times9.6$~mm$^{2}$ with $1920\times1200$~pixels (8-$\mu$m pixel pitch). The SLM imparts a 2D phase distribution $\Phi$ to the spectrally resolved wave front, which is retro-reflected back to the grating, whereupon the STWP is formed.

The SLM phase $\Phi$ is designed so as the assign to each wavelength (occupying a column of the SLM) a phase $\Phi(\lambda,x_{\mathrm{s}})$ corresponding to the spatial frequency $k_{x}(\lambda)$. We carried out the experiments with two different configurations for the phase $\Phi$. In one configuration, we split each column in two halves and encode $k_{x}(\lambda)$ and $-k_{x}(\lambda)$ in each half. This allows $\pm k_{x}(\lambda)$ to be assigned to exactly the same wavelength, but reduces the effective aperture by half [Fig.~\ref{fig:Setup}(b,c)]. Alternatively, we `interleave' the spatial frequencies $+k_{x}(\lambda)$ and $-k_{x}(\lambda)$ in neighboring columns. This approach has two advantages: (1) it helps restore the effective aperture to the full physical aperture, but introduces a slight offset between the wavelengths assigned to $\pm k_{x}$ [Fig.~\ref{fig:Setup}(d,e)]; and (2) it enables avoiding larger obstacles. It is clear from the SLM phase pattern in Fig.~\ref{fig:Setup}(c) that the transverse null cannot exceed half the height of the column dedicated to each spatial frequency (after which diffractive effects prevent the formation of a well-defined `eye' in the intensity pattern). Because both $\pm k_{x}(\lambda)$ are placed in the same column, each occupies half the SLM height, resulting in a maximum spatial width of an obstacle to be avoided of $W_{\mathrm{s}}=0.5\times0.5\times9.6=2.4$~mm (without further magnification or demagnification). Interleaving the SLM phase doubles the size of the obstacle that can be avoided to $W_{\mathrm{s}}=0.5\times9.6=4.8$~mm. However, the spectral resolution of the grating $\delta\lambda$ occupies more than one pixel on the SLM, so that this slight spectral offset does not introduce any observable impact.

We carried out experiments at three different spatial scales. In the first set of experiments, we utilize a $5\times$ demagnification system formed of two spherical lenses L$_{1}$ and L$_{2}$ of focal lengths $f=1000$~mm and $f=200$~mm, respectively, with apertures of 25.4~mm. We place a spatial filter between the two lenses in the Fourier plane between the lenses to block the spatial frequencies in the vicinity of $k_{x}=0$ (corresponding to the zeroth-order diffraction from the SLM), This arrangement with the interleaved (non-interleaved) SLM phase reduces $W_{\mathrm{ST}}$ to 1.92~mm (0.96~mm), thus enabling a maximum obstacle width to be avoided of $\sim0.96$~mm ($\sim0.48$~mm). In a second set of experiments, we utilized a $1\times$ relay system formed of a pair of identical spherical lenses L$_{3}$ and L$_{4}$ both having a focal length $f=1000$~mm and aperture 25.4~mm. Here $W_{\mathrm{ST}}=9.6$~mm for the interleaved SLM phase (maximum obstacle width $\sim48$~mm). Finally, in a third set of experiments, we utilized a $10\times$ magnification system formed of a spherical lens L$_{5}$ of focal length $f=300$~mm and aperture 50~mm, together with a spherical mirror of focal length $f=3.175$~m and diameter 317.5~mm \cite{Hall25OL}. Here $W_{\mathrm{ST}}=96$~mm for the interleaved SLM phase, allowing obstacle avoidance up to a transverse width of $\sim48$~mm.    

The SLM phase for a propagation-invariant STWP used as a base is plotted in Fig.~\ref{fig:Setup}(b), corresponding to $\theta=45.05^{\circ}$. Rather than place a physical aperture in the path of the formed STWP, we introduce a zero-phase mask in the SLM. In each SLM column (corresponding to a single wavelength $\lambda$ and its associated spatial frequency $\pm k_{x}(\lambda)$) we set the phase to zero for a finite extent equal to $5W_{\mathrm{s}}$. Because each $k_{x}=\tfrac{\omega}{c}\sin\varphi(\lambda)$ travels at a different angle $\varphi(\lambda)$ with the $z$-axis, we displace the center of the SLM zero-phase-mask vertically to lie along the curves $h(\lambda)=\pm5z_{1}\tan\varphi(\lambda)$ for each $k_{x}$, where $z_{1}$ is a constant [Fig.~\ref{fig:Setup}(c)]. The field reflected from the zero-phase-mask is eliminated by a beam block placed in the Fourier plane at $k_{x}\sim0$. After $\times5$ demagnification, a null of transverse width $W_{\mathrm{s}}$ is formed at the axial plane $z=z_{1}$. Tuning the curves $h(\lambda)=\pm5z_{1}\tan\varphi(\lambda)$ by varying $z_{1}$ changes the location at which the shadow is cast, with no need for varying a lens-based imaging system. 

\begin{figure*}[t!]
\centering
\includegraphics[width=17.6cm]{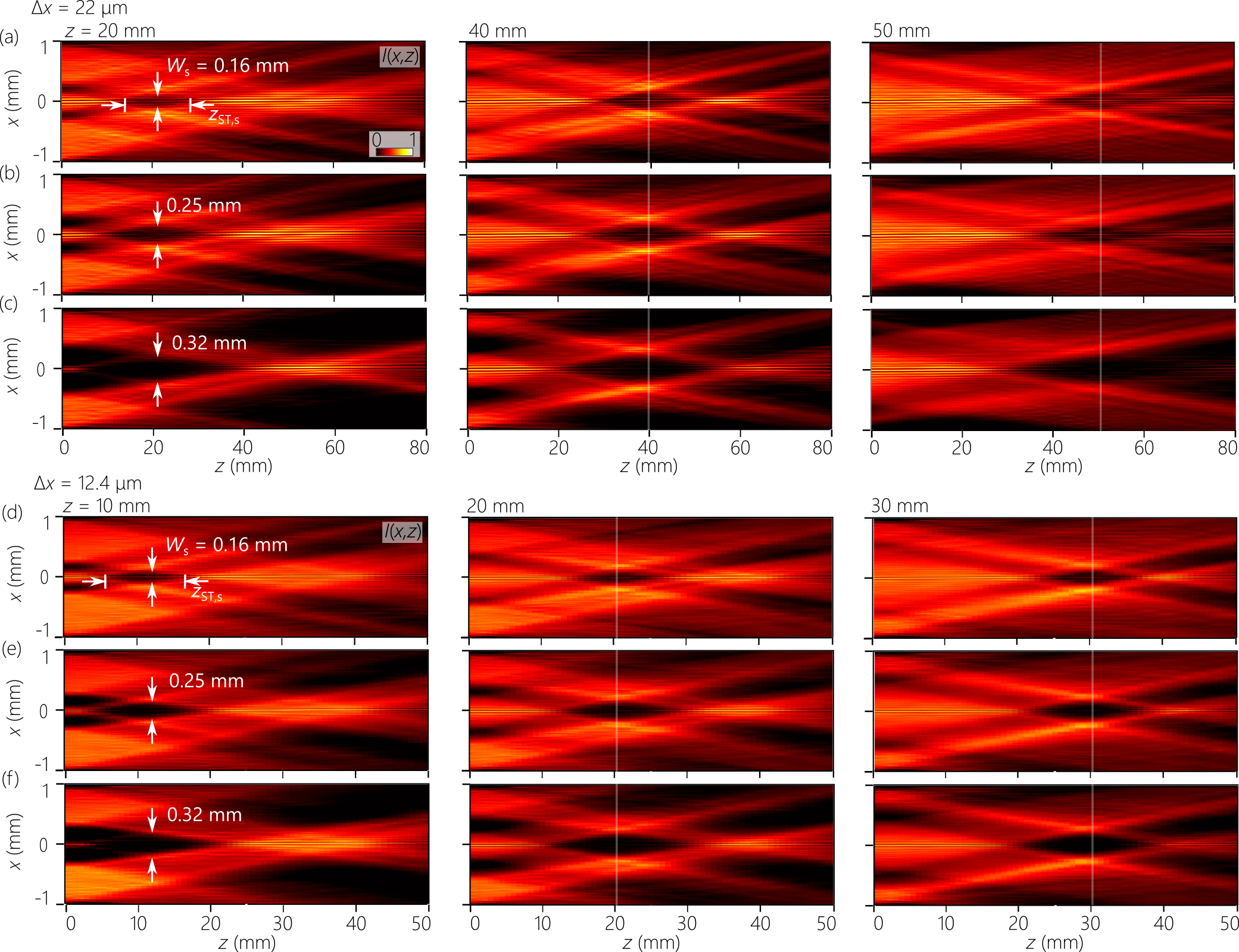}
\caption{Sub-Rayleigh shadow-projection utilizing STWPs after $\times5$ demagnification. We make use of STWPs with interleaved SLM phase distributions [Fig.~\ref{fig:Setup}(e)], $\Delta\lambda=25$~nm, and $W_{\mathrm{ST}}=1.92$~mm. (a-c) Time-averaged intensity $I(x,z)$ for an STWP with $\Delta x_{\mathrm{ST}}\approx22$~$\mu$m. The shadow is projected to the axial planes $z=20$~mm (left column), 40~mm (middle column), and 60~mm (right column). (a) $W_{\mathrm{s}}=160$~$\mu$m, (b) 250~$\mu$m, and (c) 320~$\mu$m. (d-f) Time-averaged intensity $I(x,z)$ for an STWP with $\Delta x_{\mathrm{ST}}\approx12.4$~$\mu$m. The shadow is projected to the axial planes $z=10$~mm (left column), 20~mm (middle column), and 30~mm (right column). (d) $W_{\mathrm{s}}=160$~$\mu$m, (e) 250~$\mu$m, and (f) 320~$\mu$m. The calculated intensity profile $I(x,z)$ is given in Supplementary Fig.~S1, showing excellent agreement with the measurements. The data plotted here are culled from a large collection provided in Supplementary Figs.~S2-S7, making use of both interleaved and non-interleaved SLM phase distributions.}
\label{fig:InterleavedData}
\end{figure*}

\section{Experimental results}

\subsection{Demagnification $5\times$}

We first explore sub-Rayleigh shadow-projection after implementing the $5\times$ demagnification system. A CMOS camera (DMK 33UX183) mounted on a linear translation stage scans along the $z$-axis and records the time-averaged spatial profile $I(x,z)$ of the STWP. Using \textit{non-interleaved} SLM phase distributions [Fig.~\ref{fig:Setup}(b,c)], we implemented shadow widths $W_{\mathrm{s}}=80,160,320,400$ and 480~$\mu$m, projected at axial positions $z=0,10,20,30,40,50,60,70$, and 80~mm, for STWPs with $\Delta x_{\mathrm{ST}}=12,17$, and 26.5~$\mu$m. The full set of measurements is provided in the Supplementary [Fig.~S2-S4]. Using \textit{interleaved} SLM phase distributions [Fig.~\ref{fig:Setup}(d,e)], we implemented shadow widths $W_{\mathrm{s}}=80,160,250,320,400,480,640,800$, and 960~$\mu$m, projected at axial positions $z=0,10,20,30,40,50,60$, and 80~mm, for STWPs with $\Delta x_{\mathrm{ST}}=12,22$, and 31~$\mu$m. The full set of measurements is provided in the Supplementary [Fig.~S5-S7].

\begin{figure*}[t!]
\centering
\includegraphics[width=17.2cm]{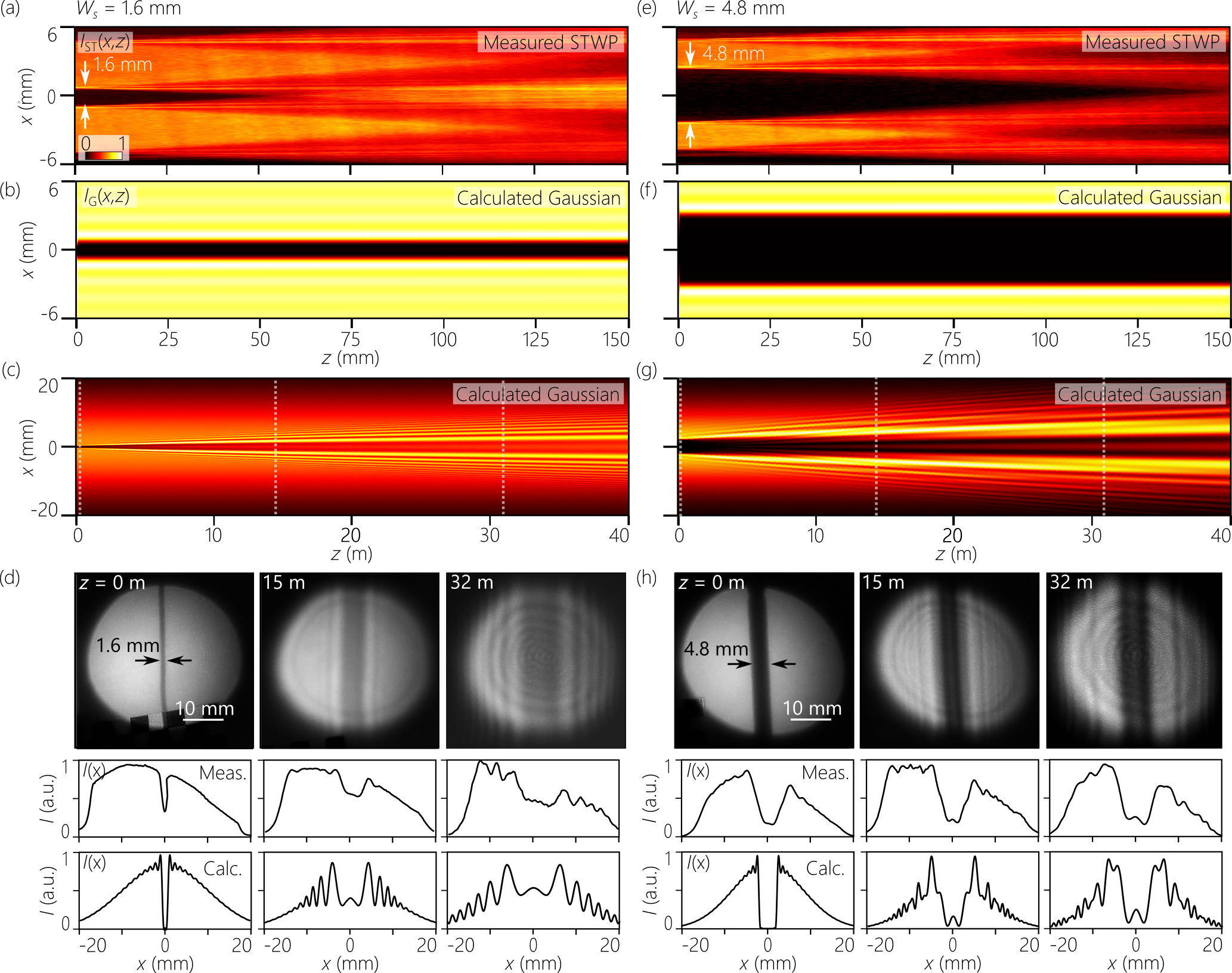}
\caption{Sub-Rayleigh-length shadow-projection utilizing an STWP after $1\times$ relay and Rayleigh-length shadow-projection utilizing a Gaussian beam. (a-e) The transverse shadow width is $W_{\mathrm{s}}=1.6$~mm and (f-j) $W_{\mathrm{s}}=4.8$~mm. In both cases, the STWP has $\Delta x_{\mathrm{ST}}=26.4$~$\mu$m, $W_{\mathrm{ST}}=9.6$~mm, $\Delta\lambda=25$~nm, and $\lambda_{\mathrm{o}}=1075$~nm. (a) The measured time-averaged intensity $I(x,z)$ for an STWP demonstrating sub-Rayleigh-length shadow-projection, with $z_{\mathrm{ST,s}}\approx60$~mm. (b) Calculated intensity $I(x,z)$ for a Gaussian beam of width $W_{\mathrm{o}}=2.56$~mm plotted over the same axial scale 150~mm as in (a); the shadow is invariant over this axial scale. The effective Rayleigh length of the shadow $z_{\mathrm{G,s}}\approx23$~m. (c) Same as (b) plotted over an axial scale of 40~m. (d) Measured intensity profiles of the transverse shadow $(x,y)$ at the axial planes $z=0,15$, and 32~m acquired down a corridor, corresponding to the dashed white lines in (c). (e) One-dimensional sections $I(x,y=0)$ through the measured transverse intensity profiles in (d), compared to the calculated profiles from (c). (f-j) Same as (a-e) for a transverse shadow of width $W_{\mathrm{s}}=4.8$~mm. The sub-Rayleigh-length for the shadow projected using the STWP is $z_{\mathrm{ST,s}}\approx150$~mm, the effective Rayleigh length for the shadow projected by the Gaussian beam is $z_{\mathrm{G,s}}\approx 23$~m. Consequently, the projected shadow is maintained along the entire length of the corridor $\sim40$~m. In (i,j) The 1D sections are taken at $z=0,15$, and 32~m.}
\label{fig:X1Data}
\end{figure*}

We plot in Fig.~\ref{fig:InterleavedData} a subset of these measurements that utilize the interleaved SLM phases. In Fig.~\ref{fig:InterleavedData}(a-c) we plot measurement results obtained after implementing an interleaved SLM phase pattern corresponding to an STWP with $\Delta x_{\mathrm{ST}}=22$~$\mu$m, with shadow widths $W_{\mathrm{s}}=160$~$\mu$m [Fig.~\ref{fig:InterleavedData}(a)], 250~$\mu$m [Fig.~\ref{fig:InterleavedData}(b)] and 320~$\mu$m [Fig.~\ref{fig:InterleavedData}(c)], each projected to 3 axial planes at $z=20,40$, and 60~mm. Projecting a shadow of large $W_{\mathrm{s}}$ close to the source or close to the end of the propagation distance $L_{\mathrm{max}}$ runs the risk of incomplete shadow formation, whereas projecting the shadow closer to the mid-range yields a complete shadow. Another data set for an STWP having $\Delta x_{\mathrm{ST}}=12.4$~$\mu$m and projecting shadows of the same widths $W_{\mathrm{s}}=160,250$, and 320~$\mu$m to axial planes at $z=10,20$, and 30~mm are plotted in Fig.~\ref{fig:InterleavedData}(d-f). The reduced $\Delta x_{\mathrm{ST}}$ reduces the length of the axially cast shadow, thereby improving the shadow formation at the beginning and end of the propagation distance of the STWP. Simulations of STWP propagation making use of the implemented SLM phases in the presence of the projected shadow are plotted in the Supplementary [Fig.~S1], which show excellent agreement with the measurements in Fig.~\ref{fig:InterleavedData}. 

\subsection{Relay $1\times$}

\begin{figure*}[t!]
\centering
\includegraphics[width=18cm]{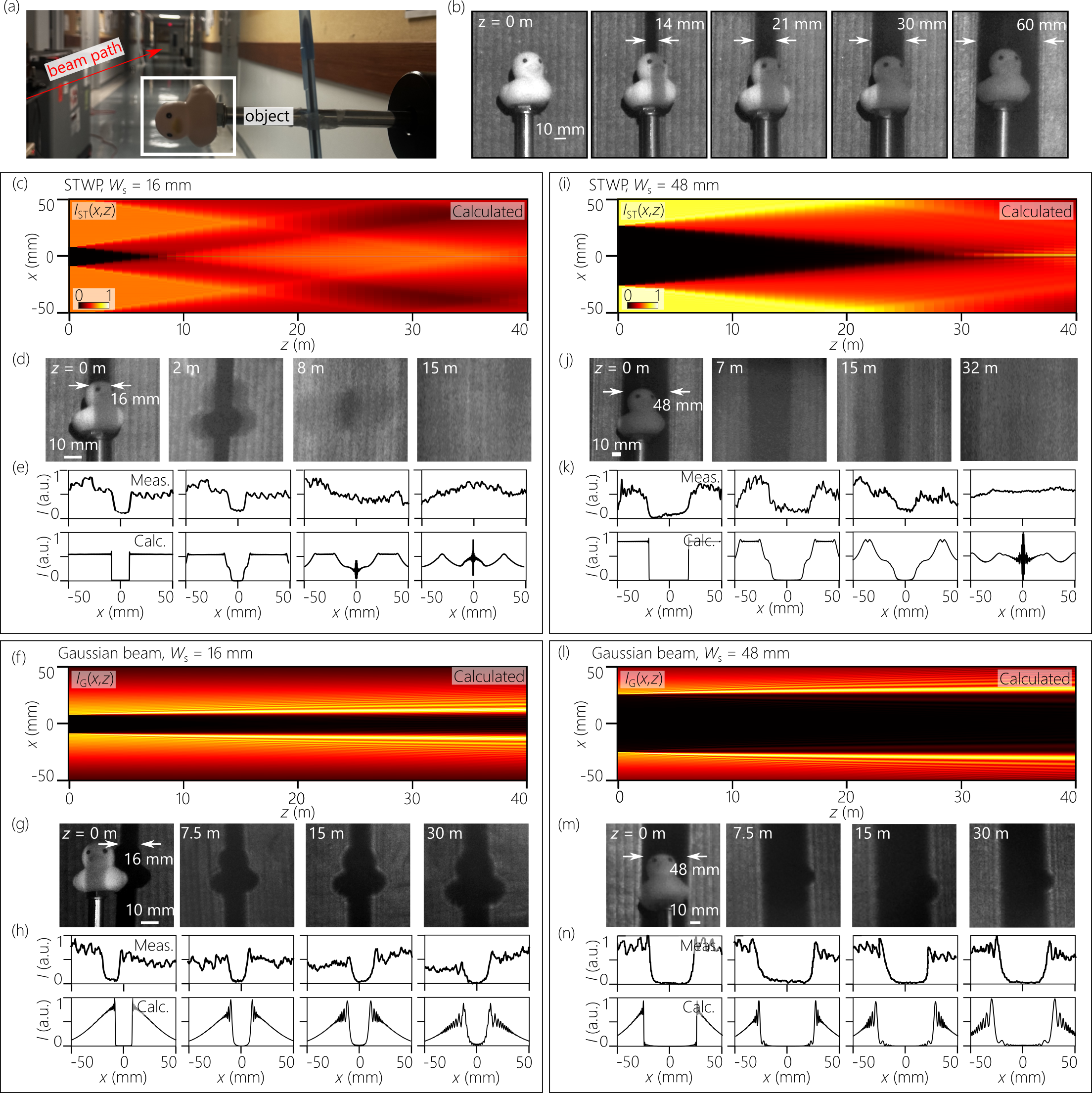}
\caption{(a) Photograph of the obstacle (a rubber duck of maximum width $\approx50$~mm) held in the corridor in which measurements are taken with $10\times$ magnification. (b) Images of the obstacle placed in the STWP LoS captured with an infrared view-finder, with projected shadows of transverse widths $W_{\mathrm{s}}=0,14,21,30$ and 60~mm. (c) Calculated time-averaged intensity distribution $I(x,z)$ for the shadow projected utilizing an STWP traveling down the corridor for 40~m. The parameters for the STWP are as follows: $\Delta x_{\mathrm{ST}}\approx0.64$~mm, $W_{\mathrm{ST}}=96$~mm, $\lambda_{\mathrm{o}}=1075$~nm, and $\Delta\lambda=25$~nm. The transverse shadow width is $W_{\mathrm{s}}=16$~mm, with expected axial length of cast shadow $z_{\mathrm{ST,s}}\approx10$~m. (d) Measured intensity profiles of the transverse shadow $(x,y)$ at the axial planes $z=0,2,8$, and 15~m, corresponding to the dashed white lines in (a). (e) One-dimensional sections $I(x,y=0)$ through the measured transverse intensity profiles in (d), compared to the calculated profiles from (c). (f-h) Same as (c-e) for a Gaussian beam of width $W_{\mathrm{G}}=300$~mm. (i-k) Same as (c-e) for a shadow width $W_{\mathrm{s}}=48$~mm. Measurements in (j,k) are taken at $z=0,7.5,15$, and 30~m. (l-n) Same as (i-k) for a Gaussian beam of width $W_{\mathrm{G}}=300$~mm.}
\label{fig:Corridor}
\end{figure*}

The $1\times$ relay system comprises a pair of spherical lenses L$_{3}$ and L$_{4}$ with $f=1000$~mm [Fig.~\ref{fig:Setup}(a)]. The unity magnification provides an effective STWP aperture of width $W_{\mathrm{ST}}=9.6$~mm, which allows us to project shadows with widths extending to $\sim5$~mm. The STWP utilized for shadow-projection has $\Delta x_{\mathrm{ST}}\approx26.4$~$\mu$m, $\lambda_{\mathrm{o}}=1075$~nm, and $\Delta\lambda=25$~nm. We first project a shadow of transverse width $W_{\mathrm{s}}=1.6$~mm. The measured time-averaged intensity $I(x,z)$ is plotted in Fig.~\ref{fig:X1Data}(a), obtained by scanning a CCD camera on the table-top for an axial distance of 150~mm. The measurement reveals an axial length for the cast shadow estimated at $z_{\mathrm{ST,s}}\approx60$~mm (the theoretical estimate is $z_{\mathrm{ST,s}}\approx42$~mm). For comparison, we plot in Fig.~\ref{fig:X1Data}(b,c) the calculated intensity distribution $I(x,z)$ when a shadow of the same width $W_{\mathrm{s}}=1.6$~mm is projected using a Gaussian beam of width 30~mm. When plotted over the same axial length of 150~mm in Fig.~\ref{fig:X1Data}(b) as done for the shadow in Fig.~\ref{fig:X1Data}(a), we find that the Gaussian shadow is unchanged. Extending the axial length monitored to 40~m, we observe a broadening of the Gaussian shadow and a gradual re-appearance of on-axis intensity (although the peak intensity does not reappear on-axis). We confirm this experimentally by performing measurements in a corridor outside the laboratory after placing the setup on a portable board. 
The shadow is cast by the Gaussian beam when a wrench with a width of 1.6~mm is placed in its path. We measured the transverse intensity profile $I(x,y)$ captured by a digital night-vision binoculars (HOTPEAK) from the beam incident on a cardboard backstop placed at axial planes $z=0,15$, and 32~m down the corridor [Fig.~\ref{fig:X1Data}(d)]. At $z=0$, the 1.6-mm-wide shadow appears clearly with a null at the center. At $z=15$~m and 32~m, the shadow broadens diffractively and low intensity replaces the null. One-dimensional sections through the measured and calculated intensities at the three axial planes are plotted in Fig.~\ref{fig:X1Data}(e), showing excellent agreement. The nominal reduction in shadow Rayleigh length is here $\eta_{\mathrm{ST,G}}\sim60$.

Similar results for a projected shadow of width $W_{\mathrm{s}}=4.8$~mm are plotted in Fig.~\ref{fig:X1Data}(f-j). The STWP-projected shadow extends to $z_{\mathrm{ST}}\approx150$~mm (theoretical estimate is $z_{\mathrm{ST}}\approx127$~mm). Minimal changes occur for the Gaussian shadow even at 40~m at the end of the corridor. The reduction in shadow Rayleigh length is here $\eta_{\mathrm{ST,G}}\sim180$.  

\subsection{Magnification $10\times$}

To accommodate even larger obstacles, we make use of the $10\times$ magnification system [Fig.~\ref{fig:Setup}(a)] comprising a spherical lens ($f=300$~mm, aperture 25~mm) and a parabolic mirror ($f=3000$~mm, aperture 300~mm). The entire setup is placed on a board that is moved out of the laboratory and into a corridor of length $\sim40$~m [Fig.~\ref{fig:Corridor}(a)]. 
 The $10\times$ magnification system increases the STWP outer width to $W_{\mathrm{ST}}=96$~mm, which allows us to consider obstacles of maximum width $\sim50$~mm. The obstacle we make use of here is a rubber duck of maximum transverse width $\approx50$~mm attached to an optical post held by a tripod positioned at the beginning of the corridor after the optical setup [Fig.~\ref{fig:Corridor}(a)]. 

We make use of an STWP with $\Delta x_{\mathrm{ST}}\approx640$~$\mu$m, $\lambda_{\mathrm{o}}=1075$~nm, and $\Delta\lambda=25$~nm. We project shadows of widths 14, 21, 30, and 60~mm using this STWP, in addition to a shadow-free STWP, onto the object, as shown in Fig.~\ref{fig:Corridor}(b). The shadow is captured with a digital night-vision binoculars (HOTPEAK) after placing a cardboard backstop. 

To demonstrate sub-Rayleigh shadow-projection, we make use of a shadow of transverse width $W_{\mathrm{s}}=16$~mm. The expected axial length of the cast shadow is $z_{\mathrm{ST,s}}\approx10$~m given the STWP parameters used, which is significantly smaller than the effective Rayleigh length of this shadow, which is $z_{\mathrm{G,s}}\approx256$~m. The calculated time-averaged intensity $I(x,z)$ is shown for the STWP projected shadow in Fig.~\ref{fig:Corridor}(c), which confirms that $z_{\mathrm{ST,s}}\approx10$~m. Measurements of the transverse projected shadow are plotted in Fig.~\ref{fig:Corridor}(d). At $z=0$ the shadow incident on the obstacle is captured. Because the projected shadow is narrower than the object, beyond $z=0$ we observe the outline of the obstacle is projected along with the shadow. Measurements along the corridor show that the shadow projected by the STWP in addition to that cast by the obstacle both rapidly heal before reaching $z=15$~m. Comparing 1D sections through the calculated intensity [Fig.~\ref{fig:Corridor}(c)] to the measured intensity profiles [Fig.~\ref{fig:Corridor}(d)] at the same axial planes show excellent agreement [Fig.~\ref{fig:Corridor}(e)].

The sub-Rayleigh shadow projected by the STWP is clear when compared to the shadow projected by a Gaussian beam of width $W_{\mathrm{o}}\approx 300$~mm. Calculations show that there is almost no change in a shadow of width $W_{\mathrm{s}}=16$~mm traveling for 40~m along the entire corridor [Fig.~\ref{fig:Corridor}(f)], which is confirmed by measurements [Fig.~\ref{fig:Corridor}(g)]. Both the projected shadow and the outline of the intercepting obstacle are maintained along the entire corridor $~\sim40$~m, which is consistent with the expected Rayleigh length of $z_{\mathrm{G,s}}\sim256$~m. The reduction in axial shadow length here is $\eta_{\mathrm{ST,G}}\sim25$.
The sub-Rayleigh shadow projected by the STWP is clear when compared to the shadow projected by a Gaussian beam of width $W_{\mathrm{o}}\approx 300$~mm. Calculations show that there is almost no change in a shadow of width $W_{\mathrm{s}}=16$~mm traveling for 40~m along the entire corridor [Fig.~\ref{fig:Corridor}(f)], which is confirmed by measurements [Fig.~\ref{fig:Corridor}(g)]. Both the projected shadow and the outline of the intercepting obstacle are maintained along the entire corridor $~\sim40$~m, which is consistent with the expected Rayleigh length of $z_{\mathrm{G,s}}\sim256$~m. The reduction in axial shadow length here is $\eta_{\mathrm{ST,G}}\sim25$.

We repeat these measurements with a projected shadow of width $W_{\mathrm{s}}=48$~mm. When projecting this shadow with the STWP, calculations [Fig.~\ref{fig:Corridor}(i)] and measurements [Fig.~\ref{fig:Corridor}(j,k)] confirm that the expected axial shadow length of $z_{\mathrm{ST,s}}\sim32$~m (before reaching the end of the corridor). On the other hand, projecting this shadow using a Gaussian beam show no change over this same distance [Fig.~\ref{fig:Corridor}(l-n)], which is consistent with the expected Rayleigh length of $z_{\mathrm{G,s}}\sim2.5$~km. The nominal reduction in axial shadow length here is $\eta_{\mathrm{ST,G}}\sim80$.


\section{Discussion}

\begin{figure}[t!]
\centering
\includegraphics[width=8.6cm]{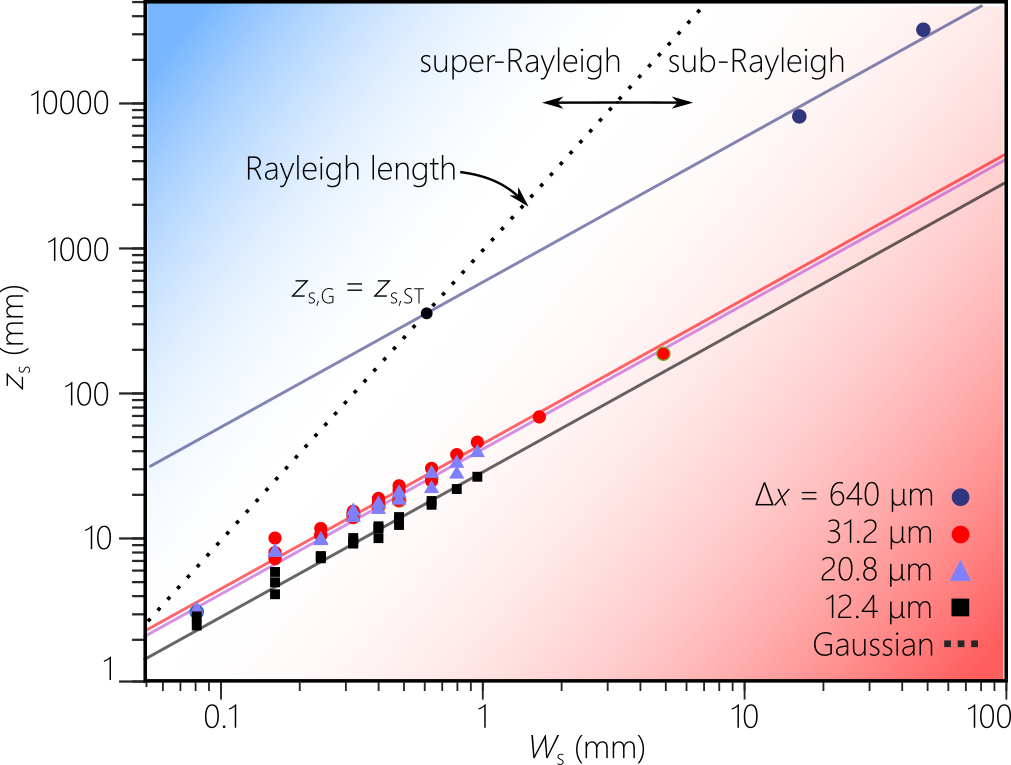}
\caption{Plot of the measured axial shadow length $z_{\mathrm{ST,s}}$ with three STWP beam FWHM sizes $\Delta x = 12$ $\mu$m (black square), $20$ $\mu$m (purple triangle), $22$ $\mu$m (red circle), and $640$ $\mu$m (dark blue circle) at various axial locations of the shadows. In addition, the shadow length of the Gaussian is plotted in a black dotted line. The regions of sub- and super-Rayleigh shadow casting is given in red and blue.}
\label{fig:ColledtedData}
\end{figure}

\textbf{Summary of results.} The space-time coupling intrinsic to STWPs has been exploited previously to \textit{increase} the propagation-invariant distance of the wave packet \cite{Bhaduri18OE,Bhaduri18OL,Hall25OE1km,Hall25OL}. Here we exploit the space-time coupling mechanism to realize an opposite trend: \textit{reducing} the Rayleigh length of a null introduced into the STWP that is relayed to a plane of interest. We collect the data reported here regarding the axial shadow length $z_{\mathrm{ST,s}}$ as a function of the transverse shadow width $W_{\mathrm{s}}$ in Fig.~\ref{fig:ColledtedData}. The data points lie along straight lines for each value of $\Delta x_{\mathrm{ST}}$, with the equation of the straight line being $z_{\mathrm{ST,s}}=W_{\mathrm{s}}\Delta x_{\mathrm{ST}}/\lambda_{\mathrm{o}}$ (which has a unity slope on the log-log scale). In addition, we plot the Rayleigh length for the shadow $z_{\mathrm{G,s}}=W_{\mathrm{s}}^{2}/\lambda_{\mathrm{o}}$ (this quadratic formula in $W_{\mathrm{s}}$ appears as a line with a slope of~2 on a log-log scale). The Gaussian line separates two domains: a sub-Rayleigh domain for the axially cast shadow below it, and a super-Rayleigh domain above it (subject to satisfying the assumptions in the formula $z_{\mathrm{ST,s}}=\tfrac{W_{\mathrm{s}}\Delta x_{\mathrm{ST}}}{\lambda_{\mathrm{o}}}$). When the line $z_{\mathrm{ST,s}}$ for fixed $\Delta x_{\mathrm{ST}}$ intersects with the Gaussian line, the axial shadow cast by the STWP is equal in length to that cast by a conventional illumination beam. All the data points lie below the Gaussian line in the sub-Rayleigh domain.

\textbf{Relationship to self-healing.} These results are related to the phenomenon of `self-healing' \cite{Bouchal99OC,Shen22JO}, which has been observed with a variety of diffraction-free beams such as Bessel \cite{Chu22EPJD} and Airy \cite{Broky08OE} beams, in addition to STWPs \cite{Kondakci18OL}. In these experiments, an opaque obstruction is placed in the beam path, whereupon a null is produced in the immediate vicinity that rapidly recovers its original transverse profile. In our work here, we do not place an obstruction in the beam path. Rather, we introduce a null at the source and then project this null to a prescribed downstream plane. The shadow formed at this plane then `heals' rapidly, so that the axial extent of the shadow is substantially smaller than that of the Rayleigh length associated with this transverse feature. This allows us to avoid the LoS obstacle while maintaining incidence on a target beyond the obstacle.

\textbf{Future developments.} The work described here can be immediately developed along multiple avenues. For example, the overall experimental strategy for LoS obstacle-avoidance can be further extended to the avoidance of multiple targets by further engineering the SLM phase distribution [Fig.~\ref{fig:Setup}(b-e)]. Moreover, although we made use of a pulsed laser in our work here, our approach is equally applicable to spectrally incoherent light, as from a superluminescent diode \cite{Yessenov19Optica} or a light-emitting diode \cite{Yessenov19OL}. In addition, other STWP structures, beyond propagation-invariant STWPs, can be used as the base wave packet to cast the LoS shadow for obstacle avoidance. Note that the optical setup used here for STWP shadow projection can be replaced by a much more compact arrangement \cite{Mhibik23OLSTWP} that relies on a new generation of rotated-chirped volume Bragg gratings \cite{Mhibik23OLBragg,Mhibik23OLRBragg}. 

\textbf{Extension to two transverse dimensions.} We made use here of only one transverse dimension $x$. Recent developments in the synthesis of STWPs localized along all dimensions \cite{Yessenov22NC,Yessenov22OL,Yessenov25NC} now allow us to extend these results to both transverse spatial coordinates. However, since the experimental synthesis of these STWPs occurs in the spatial spectral domain $(k_{x},k_{y},\omega)$ rather than the mixed domain $(x,\omega)$, it would be more straightforward to place the beam block in physical space after the spatiotemporal synthesis system and then relay it to the obstacle plane.

\textbf{Alternative illumination fields.} We have shown that the illumination field for sub-Rayleigh-length shadow projection must satisfy two desiderata: a large beam width and simultaneously a large spatial bandwidth. STWPs are not the only candidate for such a task. Another example is an optical field that with partial spatial coherence. Indeed, a field whose spatial coherence function conforms to the Gaussian-Schell model may be a candidate [Fig.~\ref{fig:CoherenceFunction}(b)], the two spatial length scales being the width of the field and the transverse spatial correlation width (or average speckle size). Another coherent-field candidate is a monochromatic Bessel beam which has two spatial length scales: the transverse beam width and the width of the central lobe.

\section{Conclusion}

We have demonstrated that utilizing STWPs as an illumination beam in shadow-projection yields a deep reduction in the effective axial Rayleigh shadow. A transverse null is projected onto the optical axis, enabling the avoidance of an on-axis obstacle. This transverse null is accompanied by an axially cast shadow that nevertheless rapidly ceases after a distance that is substantially shorter than the associated Rayleigh length. We have produced shadows of transverse widths ranging from 80~$\mu$m to 48~mm, with associated shadows that are reduced by factors of $\approx 25\times$ to $\approx250\times$ with respect to the Rayleigh length, after which the original beam profile is restored. These results may impact applications from the safe delivery of radiation therapies to non-line-of-sight optical and wireless communications.  \\

\noindent\textbf{Funding}
U.S. Office of Naval Research (ONR) N00014-20-1-2789. L.A.H. acknowledges Los Alamos National Laboratory LDRD program grant 20251140PRD1.\\

\noindent\textbf{Disclosures}
The authors declare no conflicts of interest.\\

\noindent\textbf{Data availability}
Data underlying the results presented in this paper are available upon reasonable request.


\bibliography{diffraction}

\end{document}